\documentclass[preprint,12pt]{elsarticle}

\usepackage{graphicx}

\usepackage{amssymb}
\usepackage{amsmath}

\usepackage{lineno}

\usepackage[load=prefixed]{siunitx}
\usepackage{pstricks}
\usepackage{dcolumn}
\usepackage{braket}
\usepackage{tikz}
\usetikzlibrary{patterns}
\usetikzlibrary{arrows}

\usepackage{caption}
\usepackage{subcaption}
\usepackage{multirow}
\usepackage{colonequals}
\usepackage{ mathrsfs }
\usepackage{isotope}
\usepackage{nicefrac}

\newcommand{\triumf}{{\scshape Triumf}}
\newcommand{\trinat}{{\scshape Trinat}}

\journal{New Journal of Physics}
\begin{document}

\begin{frontmatter}



\title{Precision measurement of the nuclear polarization in
  laser-cooled, optically pumped $^{37}$K}

\author[tamu-ci,tamu-phys]{B. Fenker}
\author[triumf]{J.A. Behr}
\author[tamu-ci,tamu-phys]{D. Melconian}
\author[triumf]{R.M.A. Anderson}    
\author[triumf,manitoba]{M. Anholm}
\author[telaviv]{D. Ashery}
\author[tamu-ci,tamu-chem]{R.S. Behling}
\author[telaviv]{I. Cohen}
\author[triumf]{I. Craiciu}
\author[triumf]{J.M. Donohue}     
\author[triumf]{C. Farfan} 
\author[triumf]{D. Friesen}     
\author[triumf]{A. Gorelov}
\author[triumf]{J. McNeil}
\author[tamu-ci,tamu-phys]{M. Mehlman}
\author[triumf]{H. Norton}     
\author[triumf]{K. Olchanski}
\author[triumf]{S. Smale}
\author[triumf]{O. Th\'eriault} 
\author[triumf]{A.N. Vantyghem}   
\author[triumf]{C.L. Warner}

\address[tamu-ci]{Cyclotron Institute, Texas A\&M University, 3366
  TAMU, College Station, TX 77843-3366, United States}

\address[tamu-phys]{Department of Physics and Astronomy, Texas A\&M
  University, 4242 TAMU, College Station, TX 77842-4242, United
  States} 

\address[triumf]{TRIUMF, 4004 Wesbrook Mall, Vancouver, BC V6T~2A3,
  Canada}
\address[manitoba]{Department of Physics and Astronomy, University of
  Manitoba, Winnipeg, MB R3T~2N2, Canada}


\address[telaviv]{School of Physics and Astronomy, Tel Aviv
  University, Tel Aviv, Israel}
\address[tamu-chem]{Department of Chemistry, Texas A\&M University,
  3012 TAMU, College Station, TX 77842-3012, United States}



\begin{abstract}
We report a measurement of the nuclear polarization of laser-cooled,
optically-pumped \isotope[37]{K} atoms which will allow us to
precisely measure angular correlation parameters in the
$\beta^+$-decay of the same atoms. These results will be used to test
the $V-A$ framework of the weak interaction at high precision. At the
\triumf{} Neutral Atom Trap (\trinat), a magneto-optical trap (MOT)
confines and cools neutral \isotope[37]{K} atoms and optical pumping
spin-polarizes them.  We monitor the nuclear polarization of the same
atoms that are decaying \emph{in situ} by photoionizing a small
fraction of the partially polarized atoms and then use the standard
optical Bloch equations to model their population distribution. We
obtain an average nuclear polarization of $\bar{P}=0.9913\pm0.0008$,
which is significantly more precise than previous measurements with
this technique. Since our current measurement of the $\beta$-asymmetry
has $0.2\%$ statistical uncertainty, the polarization measurement
reported here will not limit its overall uncertainty. This result also
demonstrates the capability to measure the polarization to $<0.1\%$,
allowing for a measurement of angular correlation parameters to this
level of precision, which would be competitive in searches for new
physics.

\end{abstract}


\end{frontmatter}


\section{Introduction}
\label{Intro}
Measurements in nuclear $\beta$-decay have historically contributed to
the establishment of the standard model of electroweak physics as a
theory containing massive bosons coupling only to left-handed
chirality leptons. Today, precision measurements search for and
constrain possible new physics.  For example, in isobaric analog,
mixed Fermi-Gamow Teller $\beta^{\pm}$ decays, the angular
distribution of the leptons with respect to the spin direction of the
parent nucleus is sensitive to a variety of new physics including
right-handed currents and scalar or tensor
interactions~\cite{Holstein1977, Naviliat-Cuncic1991, Herczeg2001,
  Severijns2006}.  Additionally, if we ignore this class of standard
model extensions, this measurement can be combined with other
measurements of isospin $T=1/2$ mirror-transitions to extract the
$V_{ud}$ element of the Cabibbo-Kobayashi-Maskawa quark mixing
matrix~\cite{Severijns2008, Naviliat-Cuncic2009}. This technique is
complementary to and independent of the most precise value obtained
using $T=0$ super-allowed decays~\cite{Hardy2015}. In general, to
complement high-energy searches for exotic currents in the weak
interaction, these experiments should aim for a precision of
$\sim0.1\%$~\cite{Cirigliano2013a}.

To reach this ambitious goal, we have developed the techniques at the
\triumf{} Neutral Atom Trap (\trinat{}) to confine the $\beta^+$-emitter,
\isotope[37]{K} ($I^\pi=3/2^+\rightarrow3/2^+$, $t_{1/2}=\SI{1.2}{\second}$),
in an alternating current magneto-optical trap
(AC-MOT)~\cite{Harvey2008, Anholm2014} and observe its decay
products~\cite{Melconian2007, Melconian05}.  Furthermore, the atoms
are spin-polarized by optical pumping (OP) while the MOT is off,
creating an ideal source of polarized atoms decaying nearly from rest
in an exceptionally open geometry. 

Using this setup, we have previously measured the $\nu$-asymmetry
($B_\nu$ in~\cite{Jackson1957}) to $3.6\%$
uncertainty~\cite{Melconian2007} and the $\beta$-asymmetry ($A_\beta$)
to $1.5\%$ uncertainty~\cite{Behling2015}. Although the polarization
measurement in reference~\cite{Behling2015} was consistent with tests
using naturally occurring \isotope[41]{K}, the \emph{in situ}
measurement of \isotope[37]{K} polarization was limited by
statistics. We have recently taken data for a second measurement of
$A_\beta$ with the goal of a final uncertainty less than $0.5\%$; more
precise than any previous measurement in a nucleus. Once reaching this
level of precision, we will evaluate the prospects for an
even-more-precise measurement.

\begin{figure*}[h!t]
\centering
  \includegraphics[width=0.55\textwidth,keepaspectratio]{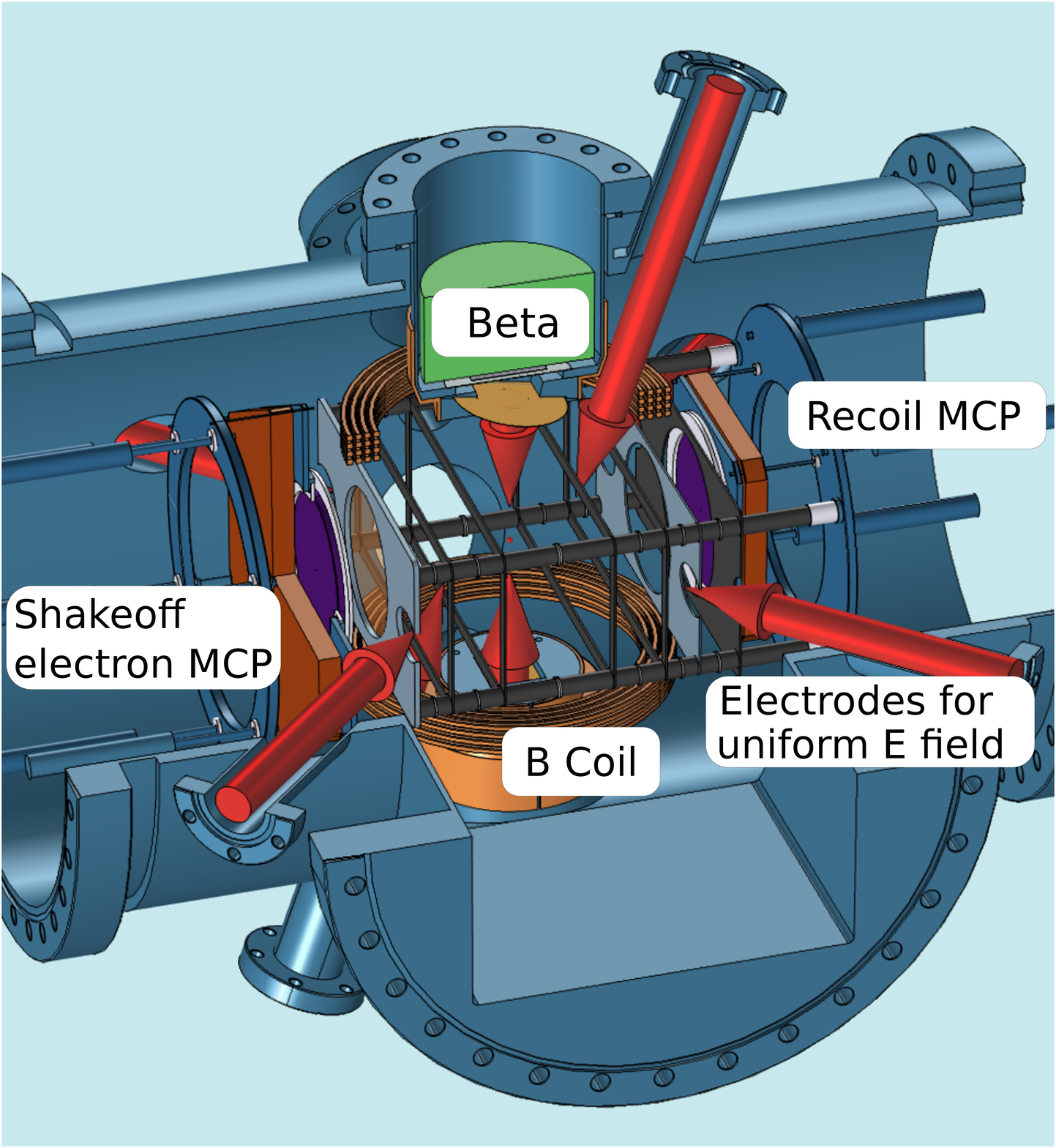}
  \resizebox{0.44\textwidth}{!}{\input{chamber3.pstex_t}}
\caption{The main \trinat{} detection chamber. The red arrows on either
  panel indicate the direction of incoming light for both the MOT and
  OP lasers.  To polarize the atoms along the axis defined by
  scintillator and silicon strip detectors, which are opaque, the
  light is brought in at a $\SI{9.5}{\degree}$ angle with respect to
  normal incidence and reflected off of a thin mirror. These detectors
  are placed symmetrically along the vertical axis and are housed in a
  re-entrant flange which is separated from the vacuum by a thin Be
  foil.  Also visible are the water-cooled magnetic field coils which
  provide the Helmholtz (OP) and anti-Helmholtz (MOT) fields as well
  as the electrostatic hoops that generate a nearly uniform electric
  field.  The recoil microchannel plate detector (MCP) is at negative
  electric potential, while the electron MCP is at positive
  potential.}
\label{fig:chamber}
\end{figure*}

In our geometry shown in figure~\ref{fig:chamber}, the
$\beta$-asymmetry can be simply determined using:

\begin{equation}
\label{eq:aobs}
A_\beta=\frac{A_\mathrm{obs}}{P}=\frac{1}{P}\frac{r^\uparrow-r^\downarrow}{r^\uparrow+r^\downarrow}.
\end{equation}

Here, $A_\mathrm{obs}$ is the observed $\beta$-asymmetry as measured in the
nuclear detectors, $P$ is the nuclear polarization, and $r^\uparrow$
($r^\downarrow$) is the rate of positrons detected along (against) the
nuclear polarization direction. In forming the asymmetry in
equation~\ref{eq:aobs}, it is possible to use a symmetric pair of
detectors along a fixed polarization axis or to use a single
$\beta$-detector and periodically reverse the sign of the
polarization. In our case, we eliminate many systematic effects by
doing both, utilizing the ``super-ratio'' technique~\cite{Gay1992,
  Plaster2012}. In addition to the nuclear measurement of $A_\mathrm{obs}$, a
measurement of $A_\beta$ requires a precision measurement of the
degree of nuclear polarization, defined by:

\begin{equation}
\label{eq:nucpol}
\vec{P}=\frac{\langle\vec{I}\rangle}{I}.
\end{equation}

Here, $\vec{I}$ is the nuclear polarization vector and, $I=3/2$ is its
magnitude. Furthermore, since the nuclear spin is greater than $1/2$,
the atoms have additional internal degrees of freedom, proportional to
the next moment of the nuclear spin projection. We define the nuclear
alignment term as:

\begin{equation}
\label{eq:alipol}
T=\frac{I(I+1)-3\langle(\vec{I}\cdot\hat{i})^2\rangle}{I(2I-1)}
\end{equation}
where $\hat{i}$ is a unit vector in the direction of
$\vec{I}$. Although it does not contribute directly to the positron
asymmetry (see footnote 7 in reference~\cite{Jackson1957}), it does contribute
to angular correlations involving the neutrino momentum, which can be
inferred from the simultaneous measurement of the momentum of the
$\beta$ and nuclear recoil. Therefore, we include it here for
completeness.

We have collected statistics for a measurement of $A_\mathrm{obs}$
with statistical uncertainty $\Delta A_\mathrm{obs}/A_\mathrm{obs} =
0.2\%$ and, therefore,
 must measure the nuclear polarization to a
similar level of precision so that the polarization measurement does
not dominate the final uncertainty.

\begin{figure}[h!t]
         \resizebox{\columnwidth}{!}{\input{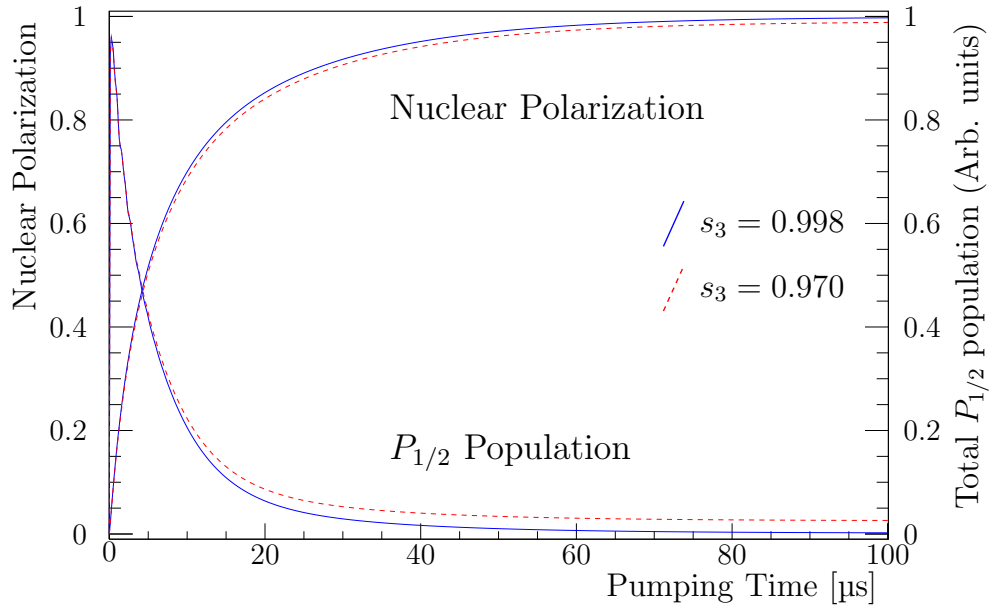}}

         \caption{Simulated time evolution of optical pumping with
           $\sigma^{\pm}$ light on the $D_1$ transition.  The
           photoionization is observed and used to infer the nuclear
           polarization by comparing to a numerical simulation of
           optical pumping. As the rate of photoionization in the
           region $t\rightarrow\infty$ decreases, the degree of
           polarization increases towards unity.  The atoms are
           considered fully polarized after the optical pumping light
           has been on for $\SI{100}{\micro\second}$ (see
           section~\ref{results}). The parameter $s_3$ gives the
           degree of circular polarization and is defined in
           section~\ref{opticalpumpinglight}. The nuclear alignment
           term follows the same measurement strategy.}
        \label{fig:principle}

\end{figure}

To polarize the atoms, we optically pump them, with the MOT off, on
the $D_1$ transition with circularly polarized ($\sigma^\pm$) light.
This accumulates atoms in the state with $m_F=\pm F$ corresponding to
complete \emph{nuclear} polarization. Here, $\vec{F}=\vec{I}+\vec{J}$
where $\vec{J}$ is the atomic angular momentum.  To monitor the
polarization, we photoionize a small fraction of the atoms which have
been excited to the $P_{1/2}$ state by the optical pumping light. This
provides a cleaner signal with fewer trapped atoms compared to
monitoring the fluorescence. The total $P_{1/2}$ population is a
sensitive probe of the nuclear polarization because atoms must have
been excited from a partially polarized $S_{1/2}$ state with
$|m_F|<F$. Therefore, the $P_{1/2}$ population is related to the
number of partially polarized or unpolarized atoms. Finally, we fit a
numerical simulation of optical pumping to the photoion time spectrum
and deduce the nuclear polarization from the result.  A typical
simulation demonstrating the principle of the technique is shown in
figure~\ref{fig:principle}. In this paper, we present the results of
this nuclear polarization measurement, which is significantly more
precise than previous results with this
method~\cite{Melconian2007,Behling2015}.

\section{Experimental Methods}
\label{expmethods}
Nuclear $\beta$-decay and atom-trap experimental methods used by
\trinat{} are described in reference~\cite{Behr2014}.  Here, we will
describe the apparatus with a particular emphasis on the polarization
measurement.  First, we give a description of the entire apparatus in
section~\ref{generalDescription}. Following this,
sections~\ref{opticalpumpinglight} and~\ref{magneticfields} describe
the two depolarizing mechanisms that lead to $|P|<1$, and
section~\ref{photoionizationlight} describes the UV light used to
monitor the nuclear polarization.

\subsection{General Description}
\label{generalDescription}
Ions of the short-lived isotope \isotope[37]{K} are delivered from
ISAC, the radioactive beam facility at \triumf, and neutralized on a
hot zirconium foil~\cite{Melconian2005}.  The atoms are then collected
in a vapor-cell MOT in a preparation chamber with $0.1\%$
efficiency~\cite{Behr1997}. To suppress a background from untrapped
atoms, they are then transferred with $75\%$ efficiency by a
red-detuned pulsed laser ``push-beam'' to a second MOT where the
precision measurement takes place~\cite{Swanson1998}. The push beam is
controlled by a double-pass acousto-optic modulator (AOM) setup, is
turned on only briefly during atom transfers, and misses the second
trap by aiming the beam $\SI{1}{\centi\meter}$ above the measurement
trap's height except during atom transfers.

Since the MOT destroys any polarization, it must be turned off and on
rapidly so that there is sufficient time to optically pump the atoms
and collect polarized decay data while the previously confined atoms
expand ballistically.  The confining forces are then turned back on to
re-collect the atoms before the cloud's expansion causes a significant
loss of atoms from the trapping region.  The trapping beam itself is
switched off to less than $10^{-4}$ of its maximum value by turning
off the first-order diffracted beam from an AOM. Any remaining trap
light is from the tail of the zeroth-order beam,
$\SI{90}{\mega\hertz}$ (15 linewidths) off-resonance. The resulting
excitation is less than $2\times10^{-4}$ of the optical pumping light.

In order to rapidly eliminate the magnetic field used for trapping, an
AC-MOT is used~\cite{Harvey2008,Anholm2014}.  In this scheme, an AC
current is run through the anti-Helmholtz coils
(see figure~\ref{fig:chamber}) instead of the usual DC current.  The
resultant magnetic field produced by the coils varies sinusoidally in
time, as does the field that results from induced eddy currents in
nearby materials, though the two components differ in phase.  Then, in
order to minimize the residual magnetic field after shutting off the
MOT, the current through the coils is shut off when the combined
magnetic field is zero.  The optimal shutoff phase is a function of
chamber geometry and material, as well as the frequency of the AC
current~\cite{Harvey2008}.

In order to trap atoms in a sinusoidally varying magnetic field, it is
necessary to vary the polarization of the MOT's trapping beam as well.
This is achieved by the use of an electro-optic modulator, set to
adjust the trapping beam between two polarization states in phase with
the magnetic field, such that a confining force is produced at all
times.

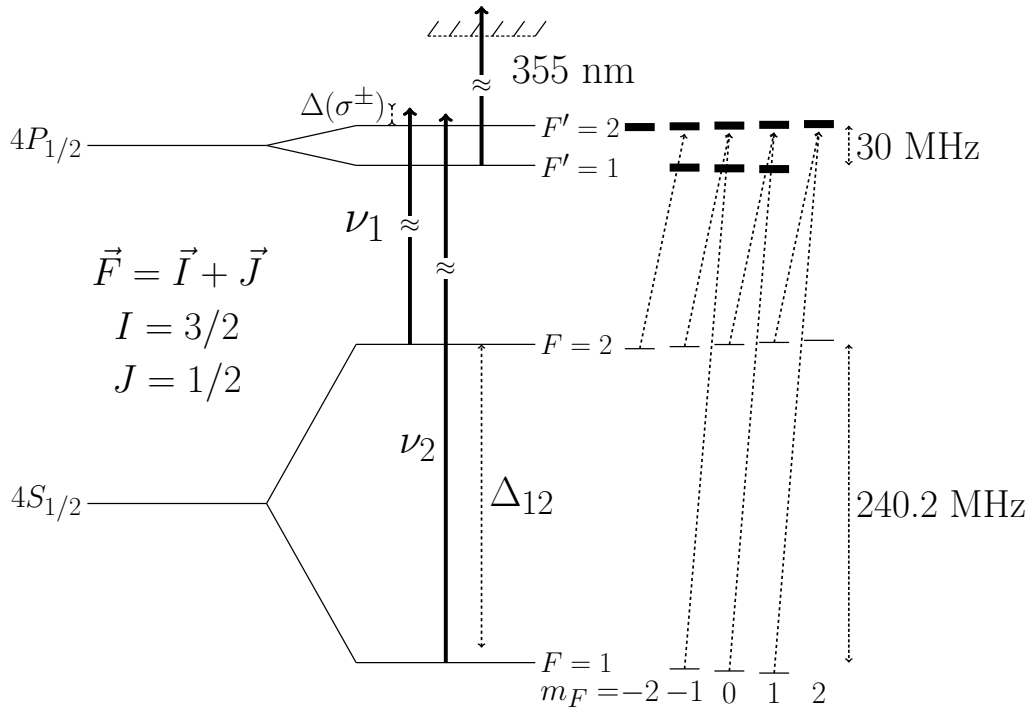
\begin{figure}[h!t]
\resizebox{\columnwidth}{!}{\begin{tikzpicture}[
    scale=0.5,
      level/.style={very thick},
      thicklevel/.style={line width=4cm, line color = red}
      trans/.style={thick,->,shorten >=2pt,shorten <=2pt,>=stealth,densely dashed},
      classical/.style={thin,double,<->,shorten >=4pt,shorten <=4pt,>=stealth}
    ]
\def\p12y{10}
\def\levellength{12};
\def\hyperfine{\levellength/2};
\def\fp2{\p12y+2*30/270*\hyperfine};
\def\fpone{\p12y-2*30/270*\hyperfine};

\draw [level] (0,\p12y) node [left,scale=1.2]{\huge$4P_{1/2}$} --(\levellength,\p12y) coordinate [right] (p12right);
\draw[level](p12right) -- (\levellength*1.5,\fp2) coordinate [right](fp2left);
\draw[level](fp2left) -- (\levellength*2.5,\fp2) node[right,scale=1.1]{\huge$F^{\prime}=2$};
\draw[level](p12right) -- (\levellength*1.5,\fpone) coordinate[right](fp1left);
\draw[level](fp1left) -- (\levellength*2.5,\fpone) node[right,scale=1.1]{\huge$F^{\prime}=1$};

\def\s12y{-14};
\def\f2{\s12y+2*240/270*\hyperfine};
\def\fone{\s12y-2*240/270*\hyperfine};
\draw[level](0, \s12y) node [left,scale=1.2] {\huge$4S_{1/2}$} -- (\levellength, \s12y) coordinate [right] (gp12right);
\draw[level](gp12right) -- (\levellength*1.5,\f2) coordinate[right] (gfp2left);
\draw[level](gfp2left) -- (\levellength*2.5,\f2) node[right,scale=1.1]{\huge$F=2$};
\draw[level](gp12right) -- (\levellength*1.5,\fone) coordinate[right](gfp1left);
\draw[level](gfp1left) -- (\levellength*2.5,\fone) node[right,scale = 1.1]{\huge$F=1$};

\node[align=center,scale=1.5] at (0.5*\levellength, -2) {\huge$\vec{F}=\vec{I}+\vec{J}$ \\ \\ \huge$I = 3/2$ \\ \\ \huge$J=1/2$};
\draw[->,line width = 0.12 cm] (1.8*\levellength,\f2) --
(1.8*\levellength,\fp2+1.2) node
[midway,left,text=black,scale=1.8,xshift=-0.25cm]{\huge$\nu_1$};
\draw[->,line width = 0.12 cm] (1.8*\levellength,\f2) --
(1.8*\levellength,\fp2+1.2) node
[midway,fill=white,draw=none,align=center]{\huge $\approx$};

\draw[->,ultra thick, line width = 0.12 cm] (2.0*\levellength,\fone) --
(2.0*\levellength,\fp2+0.8)  node [midway,yshift = -2cm,left,text=black,scale=1.6]{\huge$\nu_2$};
\draw[->,ultra thick, line width = 0.12 cm] (2.0*\levellength,\fone) --
(2.0*\levellength,\fp2+0.8)  node [midway,yshift =
4cm,fill=white,draw=none,align=center]{\huge$\approx$};

\draw[>-<,ultra thick, dashed](1.7*\levellength,\fp2)--(1.7*\levellength,\fp2+1.5) node [very near end,left,text=black,scale=1.2]{\huge$\Delta(\sigma^{\pm})$};
\draw[<->,ultra thick, dashed](2.2*\levellength,\fone+1)--(2.2*\levellength,\f2) node [midway,right,text=black,scale=1.6]{\huge$\Delta_{12}$};

\draw[->,ultra thick,  line width = 0.12 cm](2.2*\levellength,\fpone)--
(2.2*\levellength,\fp2+8) node [pos=0.6, right,xshift = 0.7 cm,text=black,scale=1.6] {\huge 355 nm};
\draw[->,ultra thick,  line width = 0.12 cm](2.2*\levellength,\fpone)--
(2.2*\levellength,\fp2+8) node [midway,fill=white,draw=none,align=center] {\huge$\approx$};

\draw[level,densely dashed] (1.9*\levellength,\fp2+6) -- (2.5*\levellength,\fp2+6);
\foreach \m in {0,...,5} {
	\draw[very thick] (1.9*\levellength + 0.12*\m*\levellength, \fp2+6) -- (1.9*\levellength + 0.06*\levellength + 0.12*\m*\levellength, \fp2+7);
}
\foreach \m in {1,...,5} {
         \def\ze{0.05*\m-0.15};
         \def\zebot{0.14*\m-0.43};
	\draw[line width = 0.27 cm] (2.75*\levellength+3*\m,\fp2+\ze)-- (2.75*\levellength + 3*\m+2,\fp2+\ze) coordinate [midway] (top);
	\draw[level] (2.75*\levellength+3*\m, \f2+\zebot) -- (2.75*\levellength + 3*\m + 2, \f2+\zebot) coordinate [midway] (bot);
}
\node[scale=1.2] at (2.75*\levellength, \fone-2.3) {\Huge $m_F=$};
\foreach \m in {-2,...,2} {
\node[scale=1.2] at ( 2.75*\levellength+3*\m+10, \fone-2){\Huge $\m$};
}
\foreach \m in {1,...,4} {
         \def\ze{0.05*\m-0.10};
         \def\zebot{0.14*\m-0.43};     
	\draw[->,dashed,ultra thick] (2.75*\levellength+3*\m + 1,\f2+\zebot)-- (2.75*\levellength+ 3*\m + 4,\fp2+\ze-0.5);
}
\foreach \n in {1,...,3} {
        \def\ze{0.05*\n-0.10};
         \def\zebot{0.14*\n-0.28};      
	\draw[line width = 0.27 cm] (2.75*\levellength + 3 + 3*\n,\fpone-\ze) --  (2.75*\levellength+ 3*\n +5,\fpone-\ze) ;
	\draw[level] (2.75*\levellength + 3 + 3*\n,\fone-\zebot) -- (2.75*\levellength+3*\n + 5,\fone-\zebot)coordinate [midway] (bot);
	\draw[->,dashed,ultra thick] (bot) -- (2.75*\levellength + 7 + 3*\n,\fp2+\ze-0.5);
}

\draw[<->,ultra thick,densely dashed](2.75*\levellength+15+3,\fp2) -- (2.75*\levellength+15+3,\fpone) node[midway,right,scale=1.5]{\huge 30 MHz};
\draw[<->,ultra thick,densely dashed](2.75*\levellength+15+3,\f2) -- (2.75*\levellength+15+3,\fone) node[midway,right,scale=1.5]{\huge 240.2 MHz};
\end{tikzpicture}}     
\caption{The fine and hyperfine structure of \isotope[37]{K} showing
  the laser transitions relevant to optical pumping.  The natural
  linewidth of the $4P_{1/2}$ state is
  $\SI{6}{\mega\hertz}$. Circularly polarized light brought in along
  the vertical axis (see figure~\ref{fig:chamber}) and tuned to the
  $D_1$ transition pumps atoms into the $F=2$, $m_F=\pm2$ state,
  resulting in a very high cloud polarization.  The parameter $\Delta$
  gives the detuning from the $F=2\rightarrow F^\prime=2$ resonance
  and is different for the $\sigma^+$/$\sigma^-$ polarization states.
  The second frequency is detuned a fixed amount, $\Delta_{12}$, from
  this frequency and optically pumps atoms which occupy $F=1$ ground
  states.  Neither $\Delta$ nor $\Delta_{12}$ are shown to scale. The
  $\SI{355}{\nano\meter}$ light continually probes the excited state
  population by photoionizing atoms from the excited $P$ states, which
  are subsequently detected by the recoil MCP.}
\label{fig:levels}
\end{figure}

Once the AC-MOT is off, the ballistically expanding atoms are
optically pumped with circularly polarized light on the $D_1$
($4S_{1/2}\rightarrow4P_{1/2}$) transition (see
figure~\ref{fig:levels}). Note that the atoms must be polarized along
the axis connecting a pair of opaque detectors as shown in
figure~\ref{fig:chamber}. In order to allow the light to propagate in
this direction, the light is brought in at a $\SI{19}{\degree}$ angle
with respect to the polarization axis and reflected off of a thin SiC
mirror before interacting with the atoms.

Furthermore, a static magnetic field, $B_z=\SI{2.3}{\gauss}$, is
applied along the quantization axis to break the degeneracy of the
Zeeman sublevels.  As a result of the optical pumping, atoms
accumulate in the 
\mbox{$4S_{1/2}|F\!=\!2,m_F\!=\!\pm2\rangle$} 
(fully stretched) state
depending on the sign of circular polarization.  This state
corresponds to complete atomic \emph{and} nuclear polarization.

To minimize systematic effects, the polarization state is reversed
every $\SI{16}{\second}$ and simultaneously a frequency shift of
$\Delta(\sigma^+)-\Delta(\sigma^-)=\SI{4}{\mega\hertz}$ (see
figure~\ref{fig:levels}) is applied. This is done in order to move
closer to the desired $m_F=\pm1\rightarrow m_{F^\prime}=\pm2$ transition
frequency while moving further from the unwanted
$m_F=\pm2\rightarrow m_{F^\prime}=\pm1$ transition, which can be excited by a
component of the optical pumping light circularly polarized with the
``wrong'' sign. Note that the sign of $B_z$ is not changed throughout
the experiment.

The nuclear polarization is measured by monitoring the total $P_{1/2}$
population of the atoms. Atoms that have been fully polarized are not
excited by the OP light and, therefore, remain in a fully stretched
$S_{1/2}$ ground state until the MOT light is switched back on. As
shown in figure~\ref{fig:principle}, observing a decrease in the
$P_{1/2}$ population implies an increase in $|P|$.

The $P_{1/2}$ population could be monitored by detecting the
fluorescence light as atoms de-excite to an $S_{1/2}$ state. However,
the collection efficiency of the first MOT as well as the flux of
\isotope[37]{K} delivered by ISAC limit the experiment to $\sim10^4$
atoms at a time. With this number of atoms, the time-resolved
fluorescence signal has a poor signal-to-noise ratio ($S/N$) and does
not provide a clean signal.

For this reason, we photoionize a small fraction of the atoms in the
$P_{1/2}$ state using UV light at $\SI{355}{\nano\meter}$ and pulsed
at a $\SI{10}{\kilo\hertz}$ repetition rate. The UV photons do not have
the energy necessary to photoionize atoms in the $S_{1/2}$ ground
state so that photoions are generated only from atoms that have been
excited to the $P_{1/2}$ state by the OP laser.

\begin{figure}[h!t]
  \includegraphics[width=\textwidth,keepaspectratio]{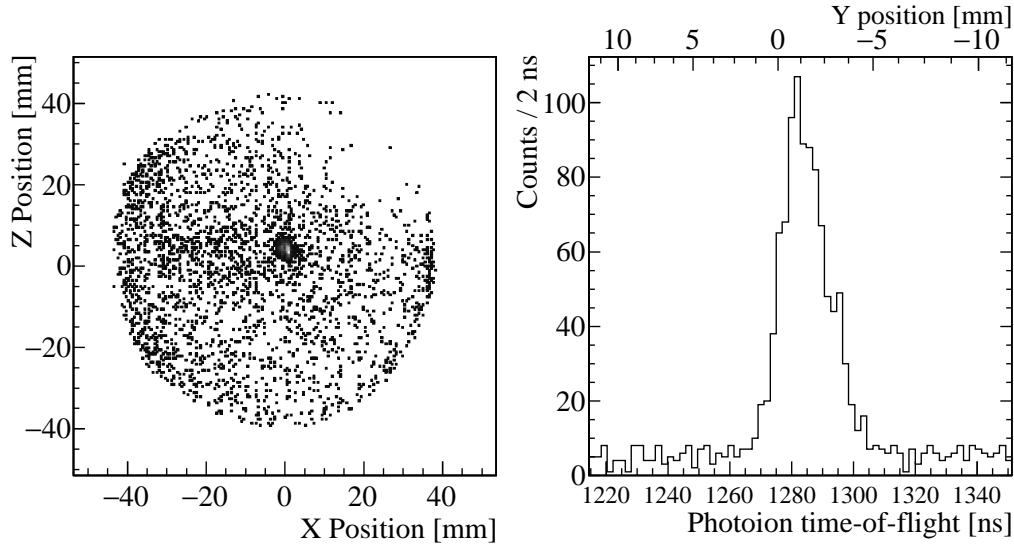}
  \caption{Photoion position and time-of-flight spectrum demonstrating
    the clean signal. The left panel shows events gated on the central
    time-of-flight peak while the right panel shows events requiring
    that the position is in the dense region at the center of the
    plate. The region of the plate nearly devoid of events has lower
    detection efficiency, but it does not affect the polarization
    measurement.}

\label{fig:photoionspectra}
\end{figure}

A uniform electric field generated by the series of electrostatic
hoops shown in figure~\ref{fig:chamber} sweeps the photoions onto the
microchannel plate (MCP) detector at negative electric potential where
they are observed in coincidence with the UV light. The MCP detector
is backed by a delay-line for position sensitivity. As a result, the
photoion spectrum shown in figure~\ref{fig:photoionspectra} is clean:
the photoions are well resolved both spatially and in time-of-flight.

However, the photoionization rate has no sensitivity to the
distribution of the partially polarized atoms throughout the Zeeman
sublevels with $|m_F|<F$.  Although this population can be made quite
small, the precision measurement described here requires knowledge of
its distribution. There have been methods developed to probe this
directly~\cite{Wang2007,Smith2006,Julsgaard2004}, but the specific
constraints of our experiment, including the relatively low number of
trapped atoms, make these impractical.  Additionally, the polarization
measurement must be non-destructive, preserving the polarization of
the atoms in order to observe the $\beta$-asymmetry in the nuclear
decay of the same atoms.

To satisfy these requirements, we adopt the method of monitoring the
$P_{1/2}$ population with photoionization as described above and
modeling the sublevel distribution of the partially polarized atoms as
presented in section~\ref{Model}.  We emphasize that the $P_{1/2}$
population, inferred from the photoionization measurement, is directly
proportional to the total partially polarized population, and the
theoretical model must only determine the sublevel distribution of
this relatively small population.

In addition to the polarization described above, a measurement of
$A_\beta$ requires a simultaneous determination of the
$\beta$-asymmetry, $A_\mathrm{obs}$. The $\beta$-asymmetry is
measured by a pair of $\beta$-telescopes placed along the vertical
polarization axis. Although this arrangement requires an extra
reflection of the OP light, it allows the measurement of
$A_\mathrm{obs}$ to have the highest sensitivity. Each
$\beta$-telescope consists of a thin Si-strip detector backed by a
thick plastic scintillator.  The scintillator fully stops the
positrons from the \isotope[37]{K} decay
($Q_{\mathrm{EC}}=\SI{6.1}{\mega\electronvolt\per\clight\squared}$) and
records their full energy.  The Si-strip detector provides position
information and, due to its low efficiency for detecting
$\gamma$-rays, suppresses the background from
$\SI{511}{\kilo\electronvolt}$ annihilation radiation.  To identify
decays that occurred within the region of optical pumping, we detect
the low energy shake-off $e^-$ by sweeping it with an electric field
towards a microchannel plate detector and observing it in coincidence
with the $\beta^+$. This combination of detectors provides an
exceptionally clean signal, almost entirely free from backgrounds.

\medskip
Having described the apparatus generally, we now give a detailed
description of the elements necessary to produce highly polarized
nuclei and measure the degree of
polarization. 

\subsection{Optical Pumping Light}
\label{opticalpumpinglight}

To obtain the highest polarization, both the $F=1$ and $F=2$ ground
states must be optically pumped. The two frequencies needed to
accomplish this are created by RF power injected directly into the
diode laser with the frequency close to the ground state hyperfine
splitting. We apply this standard technique~\cite{Kobayashi1982} at
relatively low RF power levels that produce light at about $1/2$ the
power of the carrier frequency and split from the carrier frequency by
the RF frequency.  This frequency is easily adjusted from the
hyperfine splitting of \isotope[41]{K} ($\SI{254}{\mega\hertz}$) to
\isotope[37]{K} ($\SI{240}{\mega\hertz}$) without changing the
alignment or beam spatial quality. The optical sideband strength is
monitored with a Fabry-Perot cavity and is stable in power to about
$10\%$.

\begin{figure}[h!t]
\includegraphics[width=\textwidth,keepaspectratio]{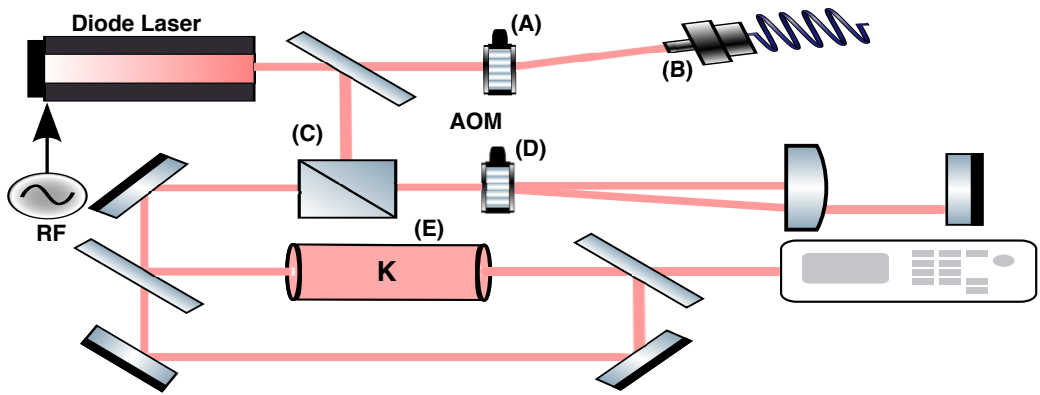}
\caption{Optical pumping light and frequency locking scheme to maintain constant
  light profile with different isotopes.  The
  $F=2\rightarrow F^\prime=2$ and $F=1\rightarrow F^\prime=1$
  frequencies are generated by RF modulation of the diode laser
  current.  The OP light is turned on and off by changing the RF input
  frequency of an AOM (A), whose first-order diffraction is steered on
  and off an optical fiber (B).  That scheme, unlike turning the RF
  power on and off, keeps the AOM at near-constant temperature,
  avoiding steering and light profile distortion as the light is
  injected into the optical fiber; thus the light power is switched
  well from zero to full value without transients.  $10\%$ of the
  light is diverted to lock the laser frequency (C).  The light is
  shifted in frequency by a tunable double-pass AOM (D) before going
  to a vapor cell of potassium (E), allowing frequency locking either
  for naturally occurring \isotope[41]{K}, or for accelerator-produced
  \isotope[37]{K}, by referencing to Doppler-free Zeeman-dithered
  saturation absorption peaks of stable isotopes~\cite{Tanaka1992}.}
\label{fig:diodelockswitch}
\end{figure}

\begin{figure}[h!t]
\includegraphics[width=\textwidth,keepaspectratio]{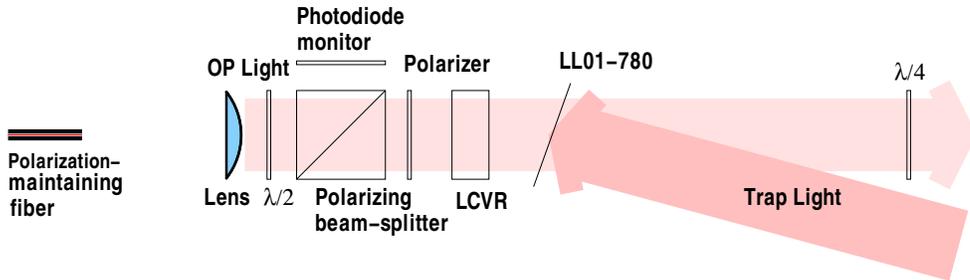}
\caption{Optical elements creating the circularly polarized $D_1$
  light. This arrangement is repeated for both OP arms. The
   liquid crystal variable retarder (LCVR) is used to control the sign
  of the circular polarization of the OP light and the LL01-780
  laser line filter is used to combine the OP and MOT
  light along the same optical path.}

\label{fig:optics}
\end{figure}

The saturation spectroscopy and double-pass AOM setup shown in
figure~\ref{fig:diodelockswitch} allows frequency locking for either
\isotope[41]{K} or \isotope[37]{K}. The light is also detuned
$\SI{1}{\mega\hertz}$ with respect to the ground-state hyperfine
splitting to completely destroy dark state coherences~\cite{Gu2003}
(see section~\ref{cpt}). Following this, the light is divided into two
beams and injected into polarization-maintaining optical fibers. The
remainder of the optical path after exiting these optical fibers is
shown in figure~\ref{fig:optics}.

After exiting the optical fiber, the OP light passes through a
polarizing beam-splitter and contrast $5\times10^4$,
$\SI{25}{\milli\meter}$ diameter suspended silver nanoparticle linear
polarizer (CODIXX ColorPol VIS 700 BC4). This is shown in
figure~\ref{fig:optics}. Next, the polarization state is determined by
the voltage applied to a liquid crystal variable retarder which either
maintains the linear polarization or rotates it $\SI{90}{\degree}$.

Since the OP and MOT light must travel the same path through the
vacuum chamber, they are combined by an angle-tuned laser line
filter. This Semrock LL01-780 nominally transmits $98\%$ of
$\SI{766.49}{\nano\meter}$ OP light while reflecting $98\%$ of the
$\SI{769.9}{\nano\meter}$ MOT light at $\SI{20}{\degree}$
incidence. The transmission of OP light changes by $4\%$ between the
linear polarization states. The output of this feeds a high-quality
$1/4$-wave plate before being injected into the vacuum chamber. Note
that there are no lenses in the path after the polarizer, avoiding
position-dependent birefringence.

The quality of circular polarization is critical to the final nuclear
polarization achieved. Any component of the light with the ``wrong''
polarization removes atoms from the fully polarized state and drives
$|P|<1$. We parametrize the quality of circular polarization with the
normalized Stokes parameter:

\begin{equation}
\label{eq:s3}
s_{3}=\frac{\mathcal{I}_+-\mathcal{I}_-}{\mathcal{I}_++\mathcal{I}_-}
\end{equation}
where $\mathcal{I}_+$ ($\mathcal{I}_-$) is the laser intensity in the $\sigma^+$
($\sigma^-$) state.

The degree of linear polarization is measured in each polarization
state along both OP arms immediately before passing through the
atom-trap viewports and $s_3$ is determined for each case. However,
stress-induced birefringence in the viewport glass can change the
light ellipticity.  We characterize this birefringence by its effect
on $s_3$ as the light passes through the viewport. If the $s_3$
parameter of the incoming light is denoted $s_3^\mathrm{in}$, then this same
parameter for the light after it has passed through the viewport is
given by~\cite{Warner2014}:

\begin{equation}
\label{eq:deltaN}
s_3^\mathrm{out}=\sin(\arcsin(s_3^\mathrm{in})+\Delta n\,kL)
\end{equation}
where $\Delta n$ parametrizes the effect of the viewport, $k$ is the
wave number of the light and $L$ is the thickness of the viewport
glass.  

We developed viewports to minimize $\Delta n$, replacing the elastomer
in a commercial viewport with PCTFE, which is compatible with
UHV~\cite{Warner2014}. We obtain $\Delta n=(-6\pm2)\times10^{-6}$ and
$(-2\pm1)\times10^{-6}$ for the two arms respectively. Although this
measurement is done with the viewports in air, we have measured the
cumulative effect of both viewports on $s_3$ both in air as well as
with the viewports under vacuum and observe no difference. This is
consistent with the pressure on the viewports having no effect on
$\Delta n$. The measured values for $s_3$ both before and after the
viewport are shown in table~\ref{tab:s3}.

\begin{table}\centering
\caption{Results of the measurement of the OP light polarization.The
  direct measurement of $s_{3}^\mathrm{in}$ is done before the
  viewport, and the value after the
 viewport ($s_3^\mathrm{out}$)
  includes a calculation of the effect of the
 birefringence in each
  viewport.}
\label{tab:s3}
   \begin{tabular}{ccD{.}{.}{4.8}D{.}{.}{5.8}}
    \hline\hline\\[-0.95em]
    & Laser port & \multicolumn{1}{c}{$s_3^\mathrm{in}$} & 
    \multicolumn{1}{c}{$s_3^\mathrm{out}$}\\
    \hline\\[-0.95em]
    & Upper & -0.9980(4) & -0.9958(8)\\[-0.65em]
    $\sigma^-$ & & & \\[-0.65em]
    & Lower & -0.9990(10) & -0.9984(13)\\[0.65em]
    & Upper & +0.9931(9) & +0.9893(14)\\[-0.65em]
    $\sigma^+$ & & & \\[-0.65em]
    & Lower & +0.9997(3) & +0.9994(5)\\
    \hline\hline
  \end{tabular}

\end{table}

After entering the vacuum chamber, the light must be reflected once as
shown in figure~\ref{fig:chamber}. The mirror used for this purpose is
coated with a commercial dielectric stack with $99.5\%$
reflectivity. We observe a change in the outgoing light's ellipticity
$|s_3^\prime-s_3|<10^{-4}$ at $\SI{9.5}{\degree}$ incidence. 

The alignment of the optical pumping light, which defines the
polarization axis, is done at the two viewports; the mirror is fixed
with mechanical precision. The result is that the light is aligned to
$\Delta\theta=\SI{1}{\milli\radian}$ with respect to the vacuum
chamber and, therefore, to the detection axis. Since the
$\beta$-asymmetry (see~\cite{Jackson1957}, equation 2) is
proportional to $\cos\theta$, this produces a negligible error of
$5\times10^{-7}$.

\subsection{Magnetic Fields}
\label{magneticfields}
A second mechanism that can drive $|P|<1$ is a magnetic field
transverse to the optical pumping axis ($B_x$) that causes Larmor
precession out of the stretched state. We have carefully designed the
apparatus to minimize eddy currents once the AC-MOT is turned off,
which in turn produce a magnetic field. Non-magnetic materials such as
316L and 316LN grade stainless steel and titanium were used wherever
possible and the chamber welds were kept thin to minimize their
magnetic permeability. We measured the relative permeability of the
welds to be $<1.25$. The vacuum chamber has a large ($\SI{12}{\inch}$)
diameter to place potentially magnetic materials as far away from the
trapped atoms as reasonably possible. The nearest material to the atoms is the
set of electrostatic hoops shown in figure~\ref{fig:chamber} which
direct the photoions onto the MCP. These are made from SIGRADUR G
grade glassy carbon, a semiconductor with resistivity
$\SI{4500}{\micro\ohm\centi\meter}$, two orders of magnitude better
than stainless steel.

In order to cancel out magnetic fields that are constant on the time
scale of optical pumping, we arranged two pairs of magnetic trim coils
exterior to the vacuum chamber. By varying the DC current in these
coils, we were able to apply a transverse magnetic field to cancel
stray fields at the location of the trapped atoms.

To optimize these settings, we optically pumped \isotope[41]{K}, which
can be trapped in large numbers and has a similar hyperfine structure
to \isotope[37]{K}. We used the same system described in this section
except that we monitored the fluorescence directly rather than the
photoionization. Keeping all the laser parameters fixed, we scanned
the trim-coil current and observed the residual fluorescence after
optical pumping. The minimum residual fluorescence corresponds to the
optimal current setting which was also used for the \isotope[37]{K}
experiment.

Additionally, the AC quadrupole magnetic field is switched off before
the optical pumping begins but induces eddy currents in the
surrounding material, which in turn produce a magnetic field. Although
the purpose of using an AC-MOT is to reduce these eddy currents by
turning off the magnetic field when it is nearly zero, we used a Hall
probe to measure an initial residual field of
$\sim\SI{103}{\milli\gauss\per\centi\meter}$, which decays to a final
value of $\sim\SI{22}{\milli\gauss\per\centi\meter}$ with a time
constant of $\sim\SI{130}{\micro\second}$. Although this measurement
was done with one vacuum flange removed, it demonstrates both the
approximate size of this effect as well as the need, described in
section~\ref{results}, to wait until this field has completely decayed
away before starting optical pumping.

\subsection{Photoionization Light}
\label{photoionizationlight}
The $\SI{355}{\nano\meter}$ UV light that photoionizes the excited
atoms is circularly polarized and has a near-$\mathrm{TEM}_{00}$ mode
with a $1/e^2$ diameter of $\SI{12}{\milli\meter}$. It comes from a
commercial diode-pumped solid-state pulsed laser making
$\SI{0.5}{\nano\second}$ pulses at $\SI{10}{\kilo\hertz}$ repetition
rate. The light propagates at $\SI{35}{\degree}$ with respect to the
optical pumping axis. After interacting with the atoms, the UV light
is reflected along the same path in order to provide a second
opportunity to interact with the atoms with $\sim90\%$ of the original
intensity. Next the sign of the polarization is reversed, and the
light again interacts with the atoms twice, although with the third
pass now at $41\%$ of the original intensity. In total, the UV light
photoionizes about $1/10^6$ atoms per pulse. The effects of the UV
light polarization on the photoionization signal are discussed in
section~\ref{Model}.

Note that the cross-section of photoionization is on the order of
$\SI{1}{\mega\barn}$, while Rayleigh scattering has a cross-section
$10^6$ lower. Therefore, the $\SI{355}{\nano\meter}$ light is
effectively a passive probe that does not disturb the system. It
either photoionizes the atom, removing it from the population so its
subsequent less-polarized $\beta$-decay is not observed, or has
negligible probability of disturbing the polarization.




\section{Theoretical Model}
\label{Model}

Having described the experimental setup, we now describe the model
used to calculate the sublevel distribution of the small fraction of
atoms that are \emph{not} fully polarized. Although this population is
small, at the current level of precision, its distribution can impact
the nuclear polarization achieved. 

Our theoretical optical pumping calculation is based on a
semi-classical approach using the density operator formalism, i.e.,
the standard optical Bloch equations with the phenomenological
spontaneous decay term $R(t)$

\begin{equation}
\label{eq:liouville}
\frac{d\rho(t)}{dt}=\frac{1}{i\hbar}[\mathcal{H}(t),\rho(t)]+R(t).
\end{equation}

We use the expressions of Tremblay and Jacques~\cite{Tremblay1990} and
extend their expressions to include the effects of two
counter-propagating beams. Because both of our frequencies come from
one laser, then are frequency shifted by an independent RF source
into two frequencies, we assume as in~\cite{Gu2003} that the
contribution of the laser linewidth to the ground-state relaxation
rate vanishes. We observed short timescale jitter of several hundred
Hertz in the RF sources and have, therefore, included a
$\SI{500}{\hertz}$ linewidth from RF sources in the ground-state
relaxation rate (see~\cite{Tremblay1990}, equation 2.37).  The external
$B$ field is included in Zeeman shifts of the magnetic sublevels.
Primarily, we consider an isotropic initial ground-state distribution,
but also consider an initial anisotropy as a systematic
uncertainty. The calculation was carried out by numerically solving
the density matrix equations, i.e., the 128 complex coupled
differential equations of the 16-level system of
figure~\ref{fig:levels}.  Additionally, an arbitrary transverse
magnetic field $B_x$, which can drive transitions with $\Delta F=0$,
$\Delta m_F=\pm1$, is included using the expressions
in~\cite{Renzoni2001}.

This model includes the two depolarizing mechanisms, discussed in
sections~\ref{opticalpumpinglight} and~\ref{magneticfields}, that lead
to $|P|<1$: ellipticity in the OP light and a transverse magnetic
field ($B_x$) which causes Larmor precession out of the stretched
state. Note that since we are pumping both ground state hyperfine
levels to prevent losses to the $F=1$ state, these are the \emph{only}
two depolarizing mechanisms. We also consider the false polarization
signal potentially produced by coherent population trapping (CPT) states 
in section~\ref{cpt}.

The transverse magnetic field and $s_3$ are highly correlated when
observing the photoionization rate: both lead to a larger fraction of
unpolarized atoms and an increase in the photoionization rate.  When
this model is fit to the experimental data as described in
section~\ref{results}, either of these mechanisms, or any combination
of them, can equally well account for the observed steady-state
photoionization and are therefore highly correlated
($>98\%$). However, the unpolarized population is distributed
differently among the $|m_F|<F$ sublevels depending on the relative
importance of the light ellipticity and the transverse magnetic field
in driving atoms out of the stretched state. Since the unpolarized
atoms make a significant contribution to the average nuclear
polarization, the relative significance of these two depolarizing
mechanisms must be considered.

In order to correctly interpret the photoionization signal as a probe
of the total $P_{1/2}$ population, we must consider the relative
photoionization cross-sections of the magnetic
sublevels. Photoionization from the $P_{1/2}$ state populates outgoing
$s$- and $d$-wave photoelectrons with the cross-section proportional
to the square of radial ($\mathcal{R}$) and angular portions of the
matrix element connecting a pair of final and initial states. Since
the angular part does not depend on the details of the central
potential, it is well known. Using a single-electron model with a
parametric central potential, Aymar, Luc-Koenig, and Combet Farnoux
calculate the total cross-section for $s$- and $d$- wave
photoelectrons and their results are
$\mathcal{R}_d/\mathcal{R}_s\approx1.7$ at
$E_\gamma=\SI{760}{\milli\electronvolt}$~\cite{Aymar1976}.

Considering the off-axis propagation as well as the multiple passes of
the UV light (see section~\ref{photoionizationlight}), the total
photoionization cross-section changes by no more than $4\%$ in our
setup compared to the assumption that all states have an equal
probability to be photoionized. The polarization results change by
$<10^{-5}$ assuming a $50\%$ uncertainty on the ratio
$\mathcal{R}_d/\mathcal{R}_s$.


\section{Coherent Population Trapping}
\label{cpt}
The multi-level system of figure~\ref{fig:levels} can support coherent
population trapping (CPT) states on three distinct sets of
$\lambda$-atomic systems ($m_F=-1,0,1$). These states are especially
problematic for this measurement as atoms in these states are not
available to be photoionized and detected, exactly mimicking our
experimental signature for good polarization, while simultaneously
having $|P| < 1$. Although CPT states are adequately described by the
model of section~\ref{Model}, we describe both how their formation is
eliminated in our setup as well as the steps that we have taken to
verify this.

First, the OP light is retroreflected such that it interacts with the
atomic cloud twice: first propagating along $+\hat{z}$ and second
along $-\hat{z}$. Since these relative velocities are different for
the two passes, the relative Doppler shift of the light frequency
between the first and second pass greatly reduces the CPT effect in
all but the coldest atoms.

To verify that they are destroyed, we performed measurements with
\isotope[41]{K}. We measure the magnitude of the CPT state similarly
to~\cite{Gu2003} by optically pumping the atoms with $\Delta_{12}$ set
to intentionally create CPT states (see
figure~\ref{fig:levels}). After the atoms are optically pumped, we
switch the frequency of the $F=1 \rightarrow F^\prime=2$ laser away
from this resonance, destroying the CPT state and allowing the atoms
that had been trapped in this state to be optically pumped to the
$m_{F}=\pm F$ state, creating a second burst of photoionization. The
relative size of the two photoionization bursts is a measurement of
the CPT fraction.

\begin{figure}[h!t]
\includegraphics[width=\columnwidth]{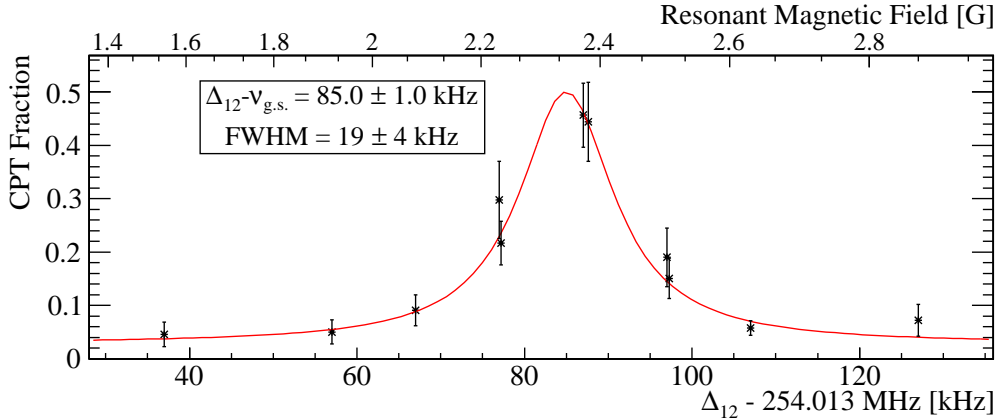}
\caption{The difference between the two OP frequencies, $\Delta_{12}$, is
  scanned near the ground state hyperfine resonance. The resonant
  frequency is $\SI{85}{\kilo\hertz}$ from the ground-state hyperfine
  splitting, providing a clean measurement of the aligned ($\hat{z}$)
  magnetic field. The width of this resonance is only
  $\SI{19}{\kilo\hertz}$ and, along with the $m_F=\pm1$ resonances, is
  carefully avoided during the \isotope[37]{K} experiment.}
\label{fig:cpt_bz}
\end{figure}

We scan the OP frequency around the $m_F=0$ ground-state hyperfine
resonance as shown in figure~\ref{fig:cpt_bz} and observe that the CPT
resonance in our system has a FWHM of only
$\SI{19\pm4}{\kilo\hertz}$. We avoid this narrow resonance, as well as
the $m_F=\pm1$ resonances, during the polarization measurement by
setting $\Delta_{12}$ to be $\SI{1.1}{\mega\hertz}$ from the
ground-state hyperfine splitting. Simultaneously, since the resonant
CPT frequency is equal to the energy difference between the two
$m_F=0$ ground states, we use this to determine the aligned magnetic
field from the second-order Zeeman shift:
$B_z=\SI{2.339(10)}{\gauss}$.

\section{Results}
\subsection{Photoionization Fits}
\label{results}
\begin{figure}[h!t]
\resizebox{\textwidth}{!}{\input{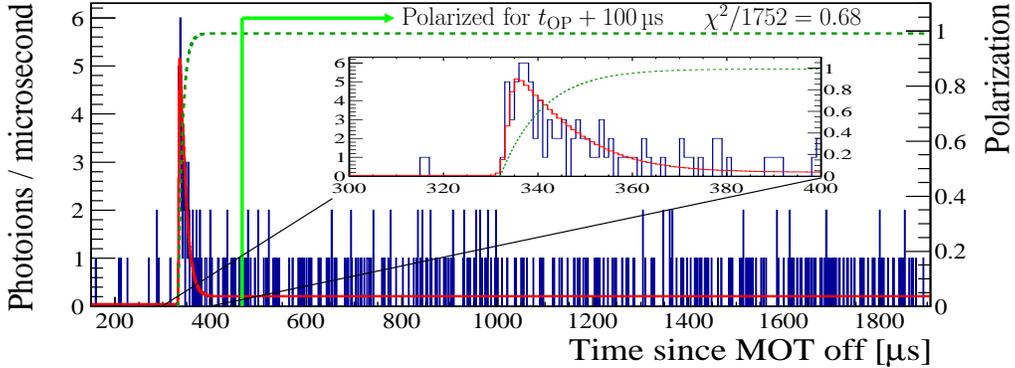}}
\caption{Result of the fit to the $\sigma^-$ polarization state with
  $t_{\mathrm{OP}}=\SI{332}{\micro\second}$ and
  $E=\SI{535}{\volt\per\centi\meter}$.  The data is shown as the blue
  histogram and overlaid with the fit result in red. The nuclear
  polarization is shown in dark green and quickly approaches one as
  atoms accumulate in the stretched state.}
\label{fig:withzoom}
\end{figure}

Figure~\ref{fig:withzoom} shows a typical photoionization curve
recorded during the experiment.  The MOT magnetic field and lasers are
switched off at $t=0$.  There was no MOT or OP light interacting with
the atoms until the OP light was turned on at
$t=t_{\mathrm{OP}}=\SI{332}{\micro\second}$. This was done in order
for the MOT magnetic field to die away as it would spoil the final
polarization as well as to give a long enough light-free region that
we use to measure backgrounds.

The atoms are fully polarized after $\SI{100}{\micro\second} $ and are
re-trapped by the MOT at $t=\SI{1906}{\micro\second}$ after expansion
from $\SI{2.0}{\milli\meter}$ to $\SI{4.5}{\milli\meter}$ FWHM.
Separate photoionization curves were recorded for the two polarization
states. This histogram is fit to the optical pumping calculation, and
the best-fit values are used to calculate the nuclear polarization and
alignment according to equations~\ref{eq:nucpol} and~\ref{eq:alipol}.

We include a constant background rate in the fitting function.  In
order to separate this background from the residual photoionization
that results from unpolarized atoms, we extend the fitting region to
begin at $t=\SI{150}{\micro\second}$, before the optical pumping has
begun.  At this point, there is no light, either from the MOT or the
OP light, illuminating the atoms. Therefore, all events between this
point and $t_\mathrm{OP}$, when the optical pumping is turned on, are
considered background.  The primary source of background events are
random coincidences between the UV pulse and the $\beta$-decay of a
\isotope[37]{K} atom, delayed by the photoion time-of-flight.
 
Also at this point, the MOT magnetic field has mostly decayed away
while still leaving enough time before the OP light is turned on to
achieve a good statistical sensitivity on the background level.  We
also observed a defect in the event timing system which caused the
recorded time to be distributed around the actual event time with a
width of $\SI{1.0}{\micro\second}$ and include this in the fit.

The variable fitting parameters were a constant background rate
described above, which is parameterized by the average signal-to-noise
ratio ($S/N$) and the OP laser intensity in each polarization state
($\mathcal{I}^{\pm}$). Additionally, the constant transverse magnetic
field ($B_x$) and one parameter describing the laser frequencies were
used as free fitting parameters.  Of these, the nuclear polarization
depends strongly only on $B_x$. The light ellipticity also strongly
influences $P$, but this is not a free fitting parameter; it is fixed
to the values of $s_3^\mathrm{out}$ shown in table~\ref{tab:s3}. Although the
transverse magnetic field is minimized in the experiment by a pair of
orthogonal magnetic field coils, its absolute value at the atoms'
position has a complicated dependence on eddy currents in the vacuum
chamber and is difficult to determine reliably.  Therefore, it is best
fit directly to the experimental data as is done here.

Other parameters, including the laser frequencies, were held constant
during the fit.  Note that both
$\Delta(\sigma^-)-\Delta(\sigma^+)=\SI{4.0}{\mega\hertz}$ and
$\Delta_{12}=\SI{239.2}{\mega\hertz}$ are well defined experimentally
and the laser linewidth is $\SI{0.2}{\mega\hertz}$. Therefore, only one
overall parameter is required to describe the laser frequencies. We
determine this overall frequency by fixing the laser intensity in the
two polarization states such that $\mathcal{I}^+=\mathcal{I}^-$ and fitting the
photoionization data to obtain the best-fit value of
$\Delta(\sigma^-)=\SI{-2.8\pm0.2}{\mega\hertz}$, which is consistent
with the direct resonance
measurement~\cite{Behr1997,Rossi2015}. Finally, the magnetic field
($B_z$) is taken from the CPT resonance measurement described in
section~\ref{cpt}.



\begin{figure}[h!t]
\includegraphics[width=\columnwidth,keepaspectratio]{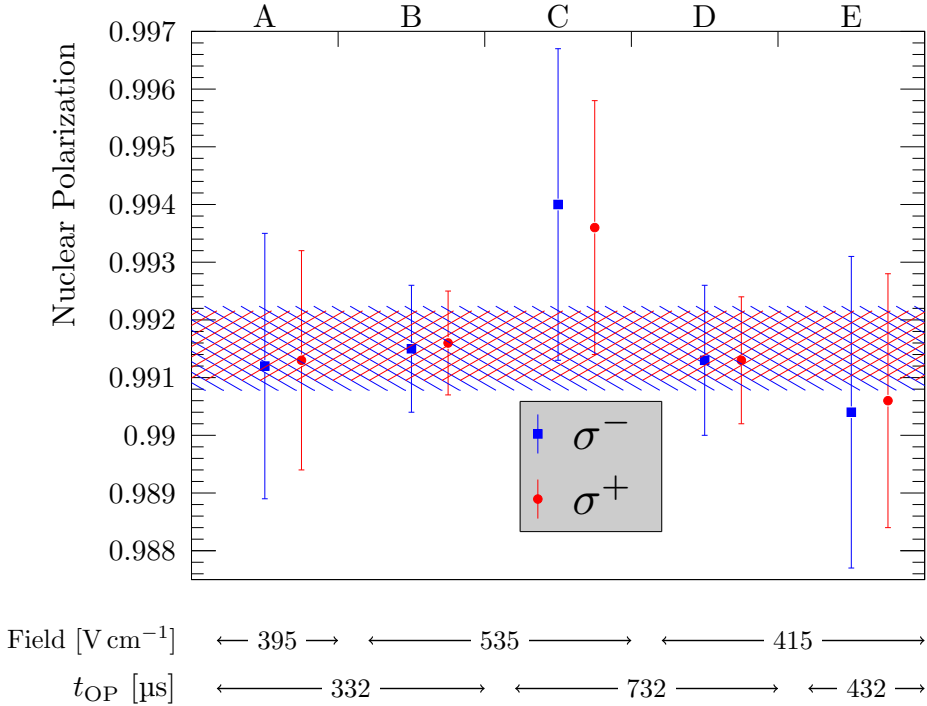}
\caption{The polarization we find as a result of fitting each set of
  data independently.  The cross-hatched region shows the $1\sigma$
  uncertainty on the polarization when combining the results of
  fitting each dataset this way. Note that the two polarization
states are not independent as the transverse magnetic field is the
same in both cases. Since there is no difference between sets, the
final result is fit to all datasets simultaneously.}
\label{fig:setbyset}
\end{figure}

Throughout the data collection, we varied the time at which we turned
on the OP light
($t_{\mathrm{OP}}=$~\SIlist[list-units=single]{332;432;732}{\micro\second})
as well as the strength of the uniform electric field to collect
photoions
(\SIlist[list-units=single]{395;415;535}{\volt\per\centi\meter}).
This resulted in five distinct datasets (not every combination was
used).  Each dataset was independently fit with the binned maximum
likelihood method, this time \emph{not} requiring that
$\mathcal{I}^+=\mathcal{I}^-$, and the results for the nuclear
polarization calculated using the best-fit parameters are shown in
figure~\ref{fig:setbyset}.  The differences in statistical sensitivity
are a result of spending different amounts of time collecting data at
the various conditions. Since there is no significant difference among
datasets, we conclude that the polarization remained constant
throughout the roughly two weeks of data taking.

\begin{figure*}[h!t]
\includegraphics[width=\textwidth,keepaspectratio]{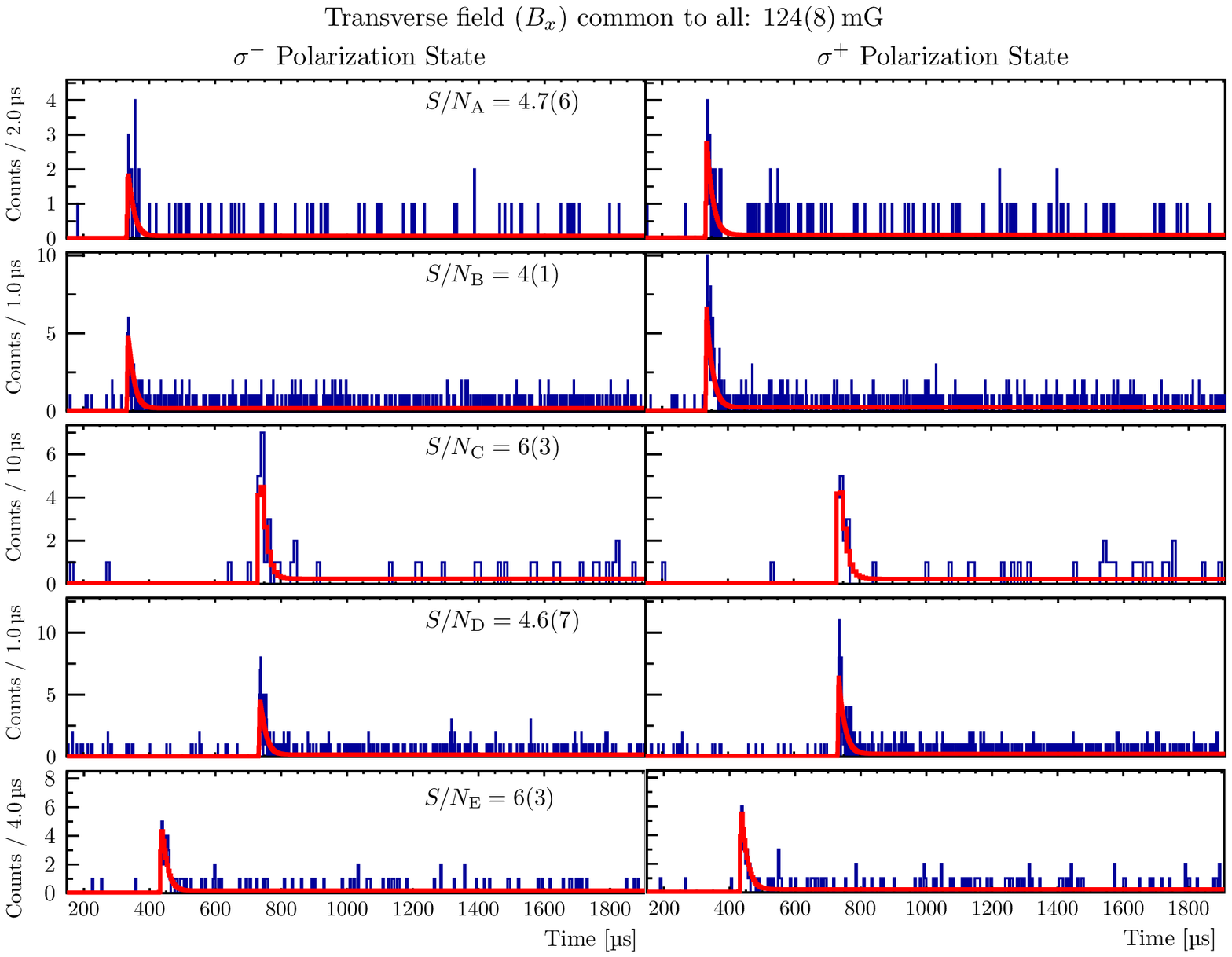}
\caption{Global fit result including a consistent set of
  parameters. The Stokes parameter, $s_3$, was fixed at its
  experimentally determined value.  A single transverse magnetic
  field, $B_x$, 
 along with separate laser intensities for each
  polarization state were fit to the entire dataset.  The
  signal-to-noise ratios ($S/N_{\mathrm{A-E}}$) were allowed to vary
  independently for each of the five datasets.  Other parameters were
  fixed as indicated in the text. The binning for each dataset was
  chosen to be as fine as possible while producing roughly equal peak
  bin contents in each set. The effects of using a uniform binning are
  discussed in section~\ref{sysuncert}. The datasets shown here from
  top to bottom correspond to the conditions shown in
  figure~\ref{fig:setbyset} from left to right.}
\label{fig:global}
\end{figure*}

Taking this into account, we performed the final analysis by fitting
each dataset simultaneously to one set of optical pumping
parameters. Since the gain of the recoil MCP detector fluctuated
throughout the run, each set was fit with an independent
signal-to-noise ratio representing a constant background in the
detector for a total of eight free fitting parameters ($\mathcal{I}^\pm$,
$B_x$, and $(S/N)_{A-E}$). The results are shown graphically in
figure~\ref{fig:global} and summarized in
table~\ref{tab:centralvalue}.

\begin{table*}\centering
  \caption{Results from the global polarization fit shown in
    figure~\ref{fig:global}.  The uncertainties listed here are purely
    statistical; the result of propagating the systematic
    uncertainties are discussed in the text. \label{tab:centralvalue}}
  \begin{tabular}{lcD{.}{.}{3.8}cD{.}{.}{3.8}}
  \hline\hline\\[-0.95em]
  \multicolumn{1}{c}{Parameter} && \multicolumn{1}{c}{$\sigma^-$} & 
  & \multicolumn{1}{c}{$\sigma^+$} \\ 
  \hline\\[-0.95em]
  Misaligned field, $B_x$ [mG]  &&  \multicolumn{3}{c}{\hspace*{-1.5em}$124(8)$}  \\
  Average $S/N$            &&  \multicolumn{3}{c}{\hspace*{0.5em}$4.7(6)$}  \\
  Laser intensity [W/m$^2$] && 
  2.33(19) && 2.26(13)\\
  Nuclear polarization     && -0.9912(7)  && +0.9913(6)  \\
  Nuclear alignment        && -0.9761(21) && +0.9770(17) \\
\hline \hline
\end{tabular}

\end{table*} 

\begin{figure}[h!t]
\resizebox{\textwidth}{!}{\input{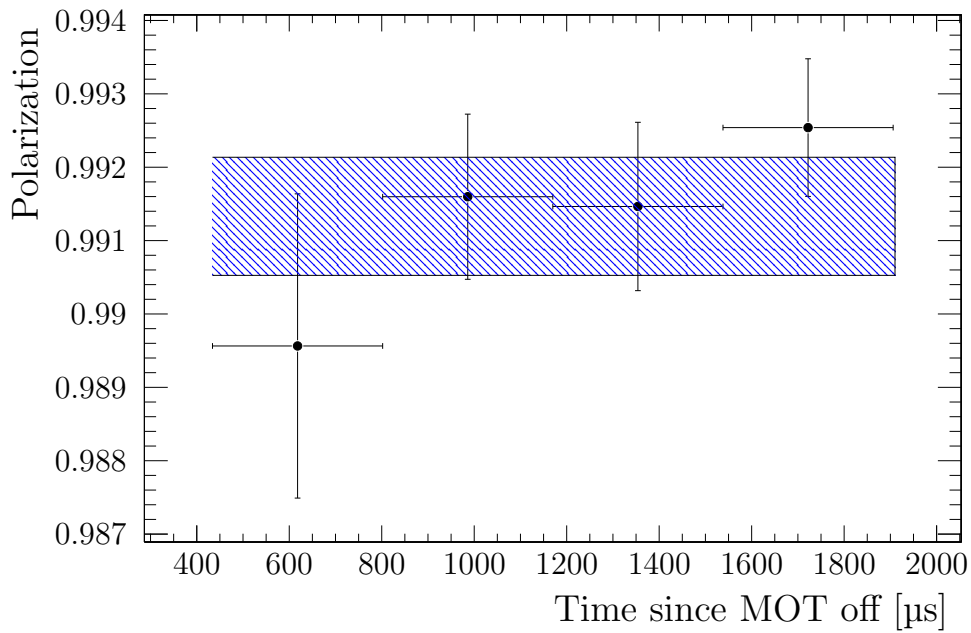}}
\caption{Time dependence of the nuclear polarization in the $\sigma^+$
  state. The shaded region shows the result with all the data
  considered while each point considers only the data in the indicated
  range in addition to the initial OP peak. The polarization is seen
  to slightly improve with time, indicating a gradual decrease in
  $B_x$.}
\label{fig:timedepend}
\end{figure}

The photoion spectra of figure~\ref{fig:global} indicate a slight
decrease in the partially polarized population even after the atoms
are considered fully polarized. This is a result of the AC-MOT
quadrupole field, and the eddy currents it creates, slowing decreasing
with time. The polarization results dividing the time when the atoms
are fully polarized into quadrants are shown in
figure~\ref{fig:timedepend}. All of the data collected with
$t_{\mathrm{OP}}=\SI{332}{\micro\second}$ is shown as this has the most
sensitivity to this effect. This figure indicates that the
polarization is improving even after $\SI{100}{\micro\second}$ of
optical pumping, although the magnitude of this effect is only
$\sim1\sigma$.  Keeping this in mind, we reiterate that the results shown
throughout this article represent the average polarization from
$t_{\mathrm{OP}}+\SI{100}{\micro\second}\rightarrow\SI{1906}{\micro\second}$.

\subsection{Systematic Uncertainties}
\label{sysuncert}
In this section, we discuss the systematic uncertainties in the
fitting routine and lay out the procedure we used to quantify them.



The final results are determined by performing a global fit to all
datasets at once.  However, it is also possible to find the weighted
average of results in figure~\ref{fig:setbyset} where each dataset is
fit independently.  The difference between these two analysis choices
gives a systematic uncertainty of $2\times10^{-4}$. Note that if we
fit each dataset independently, there are a total of twenty fitting
parameters: $\mathcal{I}^\pm$, $B_x$, and a $(S/N)$ for each of the 
five datasets.  Therefore, the global fit is preferred simply because 
it captures the same physics with fewer fitting parameters.

The uncertainty on the $s_3$ parameter is propagated to the final
result by varying the input $s_3^\mathrm{out}$ value by $\pm1\sigma$ and
comparing the results.  Although we do not expect the light's
polarization to be correlated in the two polarization states, we
conservatively treat them as though they are. This procedure gives the
most variation in the relative strengths of the two depolarizing
mechanisms, resulting in the largest difference in average nuclear
polarization.  Even with this conservative approach, the systematic
uncertainty is at most $2\times10^{-4}$ and does not limit the
measurement.

Next, the dependence of the results on the binning of the data is
studied by fitting the data with bins of width
\SIlist[list-units=single]{1;2;5;10;20}{\micro\second}. The central
value is taken from the fit with varying bin widths shown in
figure~\ref{fig:global} and we take the largest difference between any
choice of binning and this value as the systematic uncertainty.

As described in section~\ref{results}, we determine one overall
frequency by fitting the photoionization data with the requirement
that $\mathcal{I}^+=\mathcal{I}^-$. Since this requirement is only
approximately true, we relax this requirement when determining the
final results. However, we conservatively treat this condition as a
systematic uncertainty.

The magnetic field ($B_z$) has been measured by two independent
methods: the Hall probe technique described in
section~\ref{magneticfields} and the CPT field measurement described
in section~\ref{cpt}. Because the Hall probe measurement was performed
in air with one vacuum flange removed and without the presence of the
electrostatic hoops or MCP assembly, it is expected to be less
reliable than the CPT measurement. The results of these two
measurements differ by $\SI{180\pm20}{\milli\gauss}$, which is
significantly larger than the uncertainty of the CPT measurement
itself. Conservatively, this difference is treated as a systematic
uncertainty rather than propagating the smaller uncertainty on the CPT
measurement.

Finally, we allow a possible anisotropy in the initial ground-state sublevel
distribution of the atoms and characterize this by an initial
polarization $P_0$ and alignment $T_0$. We measure $P_0$ by observing
the $\beta$-asymmetry of the positrons emitted in the \isotope[37]{K}
decay before the optical pumping light is turned on. Comparing this to
the expected asymmetry ($A_\beta=-0.5706$), we conservatively measure
an initial polarization $|P_0|<0.022$. Including an initial population
distribution with this distribution produces a systematic uncertainty
of $1\times10^{-5}$.

However, $T_0$ does not produce a signal in the nuclear decay that we
can measure with the current setup.  In order to constrain this
possibility, we model the sublevel distribution of the MOT on the
$D_2$ ($F\!=\!2\rightarrow F^\prime\!=\!3$) transition. The vertically 
($\hat{z}$) propagating beams combine to produce a linearly polarized standing
wave in the $x$-$y$ plane, while the orthogonal arms produce linearly
polarized standing waves in the $x$-$z$ and $y$-$z$ planes, which
represent a combination of linearly and circularly polarized light
along the $\hat{z}$ quantization axis. Since the atom velocities
are Doppler limited, their motion averages over the polarization
gradients of the resultant electric field. Each pair of $\sigma^{\pm}$
beams have equal power and the ratio of total power propagating along
$x\,$:$\,y\,$:$\,z$ is $2\,$:$\,2\,$:$\,1$ so that the effective
ratio of linearly to circularly polarized light is $3\,$:$\,2$.  Since
the AC-MOT is deliberately turned off with $B_z$ close to zero, we
adopt the value of $B_z=\SI{100}{\milli\gauss}$.  Since a transverse
magnetic field would only serve to decrease the anisotropy, we assume
that it is zero for this calculation. The resulting population
distribution has $T_0=0.03$.  Adopting a conservative uncertainty, we
constrain the maximum initial alignment to $T_0<0.06$ and compare the
results.  These systematic uncertainties are summarized in
table~\ref{tab:systematics}.

\begin{table*}\centering
\caption{Uncertainty budget for the nuclear polarization and alignment 
  measurements. The largest systematic uncertainty arises from the potentially 
  non-zero initial alignment ($T_0$) of the atoms, which we modeled as 
  described in the text.  Also significant is the choice to perform a global
  fit rather than average the result of each dataset after a series
  of individual fits. The choice to prefer the global fit is justified
  by considering the lower number of fit parameters using this
  method.}
\label{tab:systematics}
  \begin{tabular}{llccD{.}{.}{1.1}cD{.}{.}{1.1}cccD{.}{.}{1.1}cD{.}{.}{1.1}c}
    \hline\hline\\[-0.95em]
   &&& \multicolumn{5}{c}{$\Delta P \ [\times10^{-4}]$} &\hspace*{1em} 
   &    \multicolumn{5}{c}{$\Delta T \ [\times10^{-4}]$} \\[-0.7em]
   \multicolumn{2}{l}{\hspace*{2em}Source} &&& && &&&& && &\\[-0.7em]
    &&&& \multicolumn{1}{c}{$\sigma^-$} && \multicolumn{1}{c}{$\sigma^+$} 
    &&&& \multicolumn{1}{c}{$\sigma^-$} && \multicolumn{1}{c}{$\sigma^+$} &\\
    \hline\\[-0.95em]
    \multicolumn{2}{l}{Systematics} &&& && &&&& && &\\
    &Initial alignment         &&& 3    && 3   &&&& 10  && 8   &\\
    &Global fit vs.\ average    &&& 2    && 2   &&&& 7   && 6   &\\
    &Uncertainty on $s_3^{\mathrm{out}}$ &&& 1    && 2   &&&& 11  && 5   &\\
    &Binning                   &&& 1    && 1   &&&& 4   && 3   &\\
    &Uncertainty in $B_z$      &&& 0.5  && 3   &&&& 2   && 7   &\\
    &Initial polarization      &&& 0.1  && 0.1 &&&& 0.4 && 0.4 &\\
  &Require $\mathcal{I}_+=\mathcal{I}_-$ &&& 0.1  && 0.1 &&&& 0.1 && 0.2 &\\
    \cline{5-5}\cline{7-7} \cline{11-11}\cline{13-13}\\[-0.95em]
    &Total systematic      &&& 5 && 5 &&&& 17 && 14 &\\[0.75em]
    \multicolumn{2}{l}{Statistics}
    &&& 7 && 6 &&&& 21 && 17 &\\[0.15em]
    \cline{5-5}\cline{7-7} \cline{11-11}\cline{13-13}\\[-0.95em]
    \multicolumn{2}{l}{Total uncertainty}
    &&& 8 && 8 &&&& 27 && 22 &\\
    \hline\hline
  \end{tabular}

\end{table*}

At the current level of precision, the total systematic uncertainty is
of similar, but slightly smaller, magnitude as the statistical
uncertainty.  Since the model that is fit to the experimental data
only needs to account for the small contribution to the average
polarization from the unpolarized population, all of the uncertainties
as well as the statistical uncertainty can be reduced by improving
both the light polarization and further minimizing the transverse
magnetic field to reduce the unpolarized population that must be
modeled. The final results are:

\begin{equation}
\label{eq:finalfinal}
        \begin{aligned}
        P(\sigma^+)&=+0.9913(8) \\
        P(\sigma^-)&=-0.9912(8) 
        \end{aligned}
        \qquad
        \begin{aligned}
        T(\sigma^+)&=-0.9770(22) \\
        T(\sigma^-)&=-0.9761(27)
        \end{aligned}
\end{equation}
 which represent an order of magnitude improvement compared to
 previous work~\cite{Melconian2007,Behling2015}.
\section{Discussion}


This nuclear polarization measurement is more precise than previous
measurements with the same technique and will not dominate the final
uncertainty on $A_\beta$ compared to the statistical uncertainty of
$\Delta A_\mathrm{obs}/A_\mathrm{obs} = 0.2\%$.  We note that the current
polarization measurement is limited primarily by statistics: the total
systematic uncertainty is only $5\times10^{-4}$. Therefore, we
conclude that future measurements can be sensitive enough to allow
correlation parameter measurements at the $0.1\%$ level without
significant changes to the techniques described here. In addition,
modest improvements to the apparatus will allow for an
even-more-precise measurement of the polarization in future
experiments.
 
Further increasing the light polarization and decreasing the
transverse magnetic field will both increase the average polarization
and decrease its uncertainty. With less unpolarized population to
model, the uncertainty about its distribution will lead to less
uncertainty on the nuclear polarization and alignment. Therefore, we
emphasize that improving the polarization will simultaneously improve
the precision that we can reach. Although we are continuing to
optimize the light polarization ($s_3^\mathrm{out}$), some optical elements,
particularly the liquid crystal variable retarder, preserve the
polarization better in one state than the other making it difficult to
optimize both polarization states simultaneously.  Improvements to the
trim coil system used to reduce the transverse magnetic field can also
reduce the polarization uncertainty. For example, if the magnetic
field is reduced to $1/2$ its current value and no other parameters
are changed, the statistical uncertainty is reduced by the same
factor. With careful measurements using \isotope[41]{K}, we
expect to be able to achieve this improvement.  In particular, there
is enough information from \isotope[41]{K} atoms to trim the gradient
of the magnetic field on each axis in addition to zeroing the average
value. Since we expect the systematic uncertainties to scale
similarly, it seems possible to achieve a polarization uncertainty of
$\sim0.04\%$ in upcoming measurements, allowing for an uncertainty of
$\sim0.1\%$ on the polarized correlation parameters.

\section{Conclusions}
In this paper, we have reported a precise \emph{in situ} measurement
of the nuclear polarization and alignment in optically pumped
\isotope[37]{K}. The same dataset used in these measurements contains
enough $\beta$-decay data to make a measurement of the
$\beta$-asymmetry ($A_\beta$) with an expected relative uncertainty of
$<0.5\%$. We will report these results in a future
publication. Furthermore, this work has demonstrated the capability to
measure the nuclear polarization to $<10^{-3}$, which motivates future
development towards measurements of polarized $\beta$-decay
correlations at this level of precision.

\section*{Acknowledgments}
We gratefully acknowledge the support staff of \triumf{} and ISAC and
thank S. Gensemer and A. Hatakeyama for alerting us to the effects of
CPT in optical pumping as well J. Zhang for help with weld
permeability measurements. This work was supported by the
U.S. Department of Energy under Grant No. DE-FG02-93ER40773 and Early
Career Award No. ER41747, by NSERC, by NRC through \triumf, and by the
Israel Science Foundation.






\bibliographystyle{elsarticle-num-names}
\bibliography{library_manual}







\end{document}